\documentclass[ ]{elsart}

\usepackage{graphics}

\usepackage{amssymb}
\begin{document}

\begin{frontmatter}



\title{ 
Systematic observation of tunneling field-ionization
in highly excited Rb Rydberg atoms
}


\author[ICRR]{Y. Kishimoto \thanksref{now1}},
\thanks[now1]{
Present address: Research Center for Neutrino Science, Tohoku University, 
Sendai 980-8578, Japan
} 
\author[ICRR]{M. Tada},
\author[ICRR]{K. Kominato},
\author[ICRR]{M. Shibata},
\author[SCIENCE]{S. Yamada},
\author[ICRR]{T. Haseyama},
\author[ICRR]{I. Ogawa \thanksref{now2}},
\thanks[now2]{
Present address: Department of Physics, Osaka University, Toyonaka, 
Osaka 560-0043, Japan
} 
\author[SCIENCE]{H. Funahashi},
\author[ENG]{K. Yamamoto},
\author[ICRR]{S. Matsuki \corauthref{cor1}}
\corauth[cor1]{Corresponding author.}
\ead{matsuki@carrack.kuicr.kyoto-u.ac.jp}
\address[ICRR]{
Nuclear Science Division, Institute for Chemical Research,
 Kyoto University,
 Gokasho, Uji, Kyoto 611-0011, Japan
}
\address[SCIENCE]{
Department of Physics, Kyoto University, Kita-shirakawa, Sakyo, 
Kyoto 606-8503, Japan
}
\address[ENG]{
Department of Nuclear Engineering, Kyoto University, Yoshida, Sakyo, 
Kyoto 606-8501, Japan
}

\begin{abstract}
Pulsed field ionization of 
high-$n$ (90 $\leq n \leq$ 150) manifold states in Rb Rydberg atoms 
has been investigated in high slew-rate regime.
Two peaks in the field ionization spectra were systematically observed for 
the investigated $n$ region, where the field values at the lower peak do not 
almost depend 
on the excitation energy in the manifold, while those at the higher peak 
increase with increasing excitation energy. 
The fraction of the higher peak component to the total ionization 
signals increases 
with increasing $n$, exceeding 80$\%$ at $n$ = 147.  
Characteristic behavior of the peak component and the comparison
 with theoretical predictions indicate  
that the 
higher peak component is due to the tunneling process.  The obtained 
 results show for the first time that the tunneling process 
 plays   increasingly the dominant role at such highly 
excited nonhydrogenic Rydberg atoms.        
\end{abstract}
\end{frontmatter}
%
Rydberg atoms have been utilized in fundamental physics research,
such as in the field of cavity quantum electrodynamics\cite{CavityQED}.
Area of their application fields has recently been  more
enhanced with the development of the quantum computing and its
related quantum measurements\cite{Qcmp}. 
In these experimental studies,
the field ionization method has been
widely used 
for the detection of highly excited atoms.
Most applications of Rydberg states up to now
have been performed by using  alkali atoms with the principal
quantum number $n$ less than 60 or so and it is 
known in this region that the field ionization  occurs mostly around the 
classical saddle-point field-value given by 
$F_c = W^{2}/4$ (in a.u.), where $W$ is the binding energy of the Rydberg 
states. There often observed are two peaks in 
the field ionization spectra due to the adiabatic and  
non-adiabatic processes in the time evolution of the Rydberg atoms in 
the electric field\cite{Stebbings,Gallagher}.  This ionization is 
due to the level mixing in the ionic core 
of the bound blue state with the continuum red states coming down from 
the higher $n$ states(autoionization-like 
process)~\cite{Kleppner1976,Stebbings,Gallagher}. 
This field ionization due to the level mixing effect occurs at the 
field region between the classical saddle point value and the   
value expected from the tunneling process\cite{Stebbings,Gallagher}. 
At the low $n$ (less than $\sim$ 60) region,  the ionization due to 
the tunneling process has 
not been observed usually in nonhydrogenic atoms because the ionization 
due to the 
autoionization-like process has much shorter ionization-lifetime   
than that from the tunneling process (see, however, Neijzen and 
D\"onszelmann\cite{Neijzen} and Rolfes, Smith and 
MacAdam\cite{Rolfes}).    

Since the level mixing in the ionic core interior is  weaker with 
increasing $n$ by a factor of $n^2$,  it is naturally 
expected that the competition between the autoionization-like 
and the tunneling processes occurs at some $n$ region and thereafter 
the tunneling process will dominate the ionization process.  
To our best knowledge, 
 no detailed investigations have been reported for the field 
ionization behavior in highly excited Rydberg atoms with $n$ higher 
than 70.  In addition to the above intrinsic interests in the physics of 
pulsed field ionization mechanism in highly excited Rydberg atoms, 
such studies are 
inevitably important to extend the applications of the Rydberg atoms 
to still higher excited region where the coupling to the external 
electromagnetic field is increasingly stronger and thus open new areas  
of application fields.            

We report here the experimental results of the detailed systematic 
investigation of field ionization behavior of the highly excited Rb 
Rydberg states with $n$ ranging from 90 to 150. Specifically we 
focussed on the high-$n$ manifold states which are degenerate in 
energy at zero electric field except for some low angular-momentum 
($\ell$) states, but are split under the 
electric field.  After exciting these splitting manifold states from a    
low-lying 5$p_{3/2}$ state with lasers under a small electric field, 
pulsed electric field was applied to ionize the states in high slew 
rate regime. It was 
found for the first time that the tunneling process is increasingly 
the dominant process in the pulsed field ionization at such highly excited 
Rydberg atoms.  Here the systematic occurrence of the tunneling 
process and comparison of the results with theoretical predictions 
are presented and discussed with  
applications of such highly excited Rydberg states.   

\begin{figure}
\begin{center}
	\resizebox{60mm}{!}{
		\includegraphics{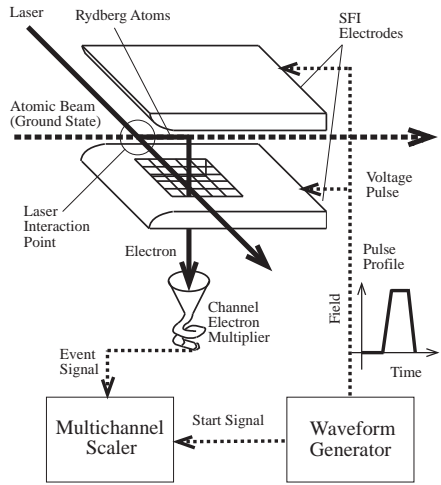}
	}
	\caption{
	Schematic diagram of the present experimental setup. 
	}
	\label{fig:SFIbox}
\end{center}
\end{figure}
Schematic diagram of the present experimental setup 
is shown in Fig.\ \ref{fig:SFIbox}.
A thermal atomic beam of Rb was introduced at the center of the  
field ionization electrodes, where the $^{85}$Rb atoms in the ground 
state $5s_{1/2}$ were excited to the Rydberg manifold states 
through the second excited $5p_{3/2}$ state 
by the two-step laser excitation.
A diode laser (780\,nm) 
and a dye laser (479\,nm) with a dye of coumarin 102 pumped by a 
Kr ion laser 
were used for the first and second step excitations, respectively. 
The polarizations of the laser lights adopted in the present experiment 
are 
parallel and perpendicular to the electric field applied for the first and 
the second lasers, respectively\footnote{
No appreciable change in the 
experimental results was observed, however, with  other 
configurations such as both parallel and perpendicular cases. 
}.

A linearly ramped electric field was applied for the field ionization in the 
present experiment. For this purpose a sequence of voltage pulse was generated
by a waveform generator AWG420 (Sony Tektronix)
and amplified by fast amplifiers before applying to the electrodes.
The electrodes consist of two parallel plates 
of 52\,mm length and 40\,mm width,
the distance between them being 24\,mm.

The electrons liberated by the field ionization were
guided to a channel electron multiplier
through two fine-mesh grids  placed in one of the electrode plates.
Ionization signals of electrons from the channel electron multiplier
were amplified by a preamplifier and a main amplifier
and then fed to a multichannel scaler P7886 (FAST ComTek)
after the pulse height discrimination.
The ionization events were counted as a function of the elapsed time
from the starting time of the ramp field with the dwell time of 500ps.
From the correspondence of the applied electric field and the time 
bin, the observed timing spectra were converted into the field 
ionization spectra as a function of the applied electric field. 
%
\begin{figure}
\begin{center}
	\resizebox{50mm}{!}{
		\includegraphics{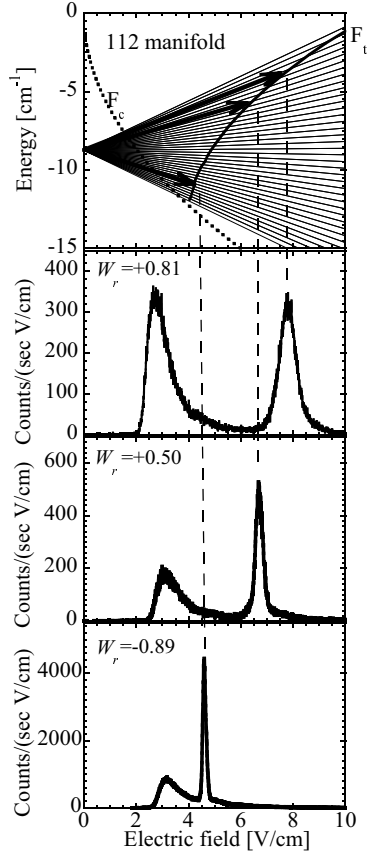}
	}
	\caption{
Relevant Stark energy diagram and field ionization spectra for the 112 
manifold states. Only 
every one of three states in the split $n$=112 Stark manifold states are 
shown for clarity.  Dotted and solid lines are the theoretical 
predictions of the ionization threshold values expected from the classical 
saddle point 
ionization ($F_{\rm c}$) and from the tunneling process ($F_{\rm 
t}$), respectively.     
The manifold states were initially populated 
under the static electric field of 75 mV/cm, and subsequently ionized 
with a pulsed electric field at the slew rate of 13.4 V/(cm$\cdot \mu$s). 
Here $W_r$ is a parameter representing the excitation position of 
the manifold states in which the values +1.0 and -1.0  correspond to the uppermost 
(highest energy) and the lowermost (lowest energy) states, respectively.     
}
	\label{fig:112m_tunneling}
\end{center}
\end{figure}
%

The manifold states were excited under a small electric field applied. 
The applied field was set to $\sim 0.7F_{e}$, where 
$F_{e} (= 1 / {3n^5})$ is the field at which $n$ and $n \pm 1$ 
manifolds cross first. 
This field is larger than the 
first avoided-crossing field between the $p$ state and the adjacent manifold 
in $^{85}$Rb, 
and yet 
smaller than $F_{e}$.  It is noted here that the manifold states can 
be excited from the low-lying $p$ state only under the electric field 
because they are intrinsically  
high $\ell$ states in zero field.  Even under the electric field, 
however, the 
extreme states near the bluest and the reddest states are hard to be 
excited appreciably, thus being unable to be used as 
a reference to the excitation position of the manifold states.  Precise 
position of 
the excited states in the manifold was therefore determined from the 
wavelength difference of the second laser relative to 
the excitation of $s$ and $p$ states at the applied electric field.  
For this purpose, exact 
excitation energies of the $s$ and $p$ states, which already enter into 
the manifold and thus not being well separated from the manifold states, 
were evaluated from the calculated 
excitation energies at which the $s$ and $d$ components in their wave functions 
have maxima among the close-lying manifold states.  The Stark energy 
structure calculated with the method of 
Hamiltonian diagonalization\cite{Kleppner1979} is shown to be precise 
enough for this purpose even for such high-lying states\cite{Kishimoto}.  

Typical field ionization spectra measured at the $n=112$ manifold with 
slew rate 13.4V/(cm$\cdot \mu$s) are shown in Fig.~\ref{fig:112m_tunneling}.  
 There  
observed are two prominent peaks, the higher one of which moves appreciably 
in its field ionization value  
with the excitation position in the manifold, while the lower one of which 
shows almost no dependence on the excitation position in the manifold.  

\begin{figure}[h]
\begin{center}
	\resizebox{50mm}{!}{
		\includegraphics{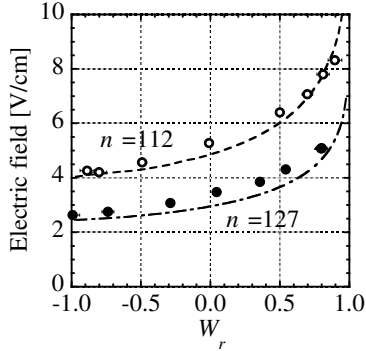}
	}
	\caption{
	Observed values of the ionization field  
	corresponding to the higher peak component as a function of the 
	excitation position in the manifold for $n$ = 112 and 127.  
	Lines are the 
	theoretical predictions of the field values expected from 
	the tunneling process. 
	}	
	\label{fig:112-127}
\end{center}
\end{figure}

Observed values of the ionization field corresponding to the higher 
peak are plotted  in Fig.~\ref{fig:112-127} for the two cases of $n$ 
as a function of the excitation position $W_r$ in the manifold.  
Here the excitation position $W_r$ in the 
manifold with its binding energy $W$ is expressed by

$W_{r} = 2\frac{W-W_{\rm min}}{W_{\rm max}-W_{\rm min}} -1$,

where $W_{\rm max}$ and $W_{\rm min}$ are the binding energies of the uppermost 
(bluest state) and the 
lowermost (reddest state) energy states in the manifold, respectively. 

Also shown in this figure with dotted and dashed lines are the theoretical 
predictions of 
the ionization field from the tunneling process; the expected 
ionization field 
from this process was calculated with a semi-empirical formula of 
ionization rate in the electric field by Damburg and 
Kolosov\cite{Damburg}. Although this formula was originally developed  
for  Hydrogen and less 
accurate in its prediction of ionization lifetime compared to the 
density-of-state method by Luc-Koenig and 
Bachelier\cite{Luc-Koenig}, Damburg and Kolosov prediction for the 
ionization field is considered to be accurate enough for the present 
purpose.  

%
%
The 
theoretical predictions  are in good agreement with the experimental 
results if the ionization paths under the applied pulsed electric field 
from the respective excitation 
positions are along the diabatic trajectories in the Stark energy 
diagram as 
shown by thick solid lines with arrows in Fig.~\ref{fig:112m_tunneling}.  
This agreement was found to be  well satisfied for all the 
investigated Rydberg states with 90 $\leq n \leq$ 150.  

\begin{figure}
\begin{center}
	\resizebox{60mm}{!}{
		\includegraphics{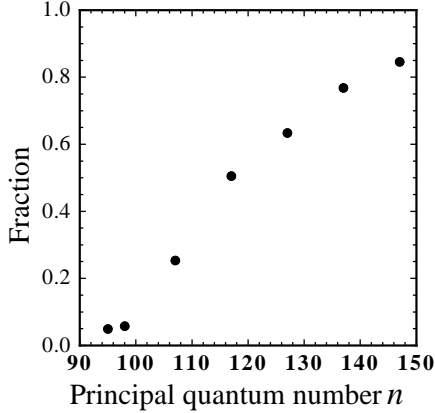}
}
	\caption{
	Fraction of the higher peak component 
	to the total ionization signals as a function of $n$.
	Excitation position in the manifold is at 
	the center of the manifold, 
	$W_r = 0$, where the external electric field 
	$F\sim 114 \cdot (100/n)^5$\,mV/cm was applied 
	at its initial excitation.  Slew rate of the applied 
	pulse is 13.4\,V/(cm$\cdot \mu$s). 
	}
	\label{fig:fraction_n} 
\end{center}
\end{figure}

The fraction of the higher peak component to the total ionization 
signals is shown in Fig.\ref{fig:fraction_n} as a function of $n$. 
The fraction increases 
with increasing $n$, exceeding  80 \% at $n$ = 147. This trend 
is consistent with the theoretical expectation that 
the level-mixing (autoionization-like process) is less 
effective with increasing $n$. 

\begin{figure}
\begin{center}
	\resizebox{60mm}{!}{
		\includegraphics{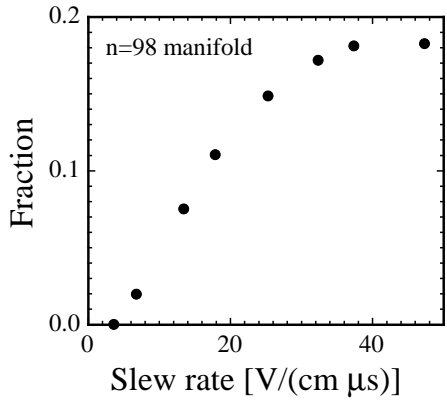}
}
	\caption{
	Slew rate dependence of the fraction of the higher peak component 
	to the total ionization signals for the $n$=98 manifold. 
	}
	\label{fig:slewrate} 
\end{center}
\end{figure}

Another important experimental result is that this fraction of 
tunneling process increases 
with higher slew rate as shown in Fig.~\ref{fig:slewrate}. This 
feature is also consistent with the theoretical prediction  
that the field ionization due to the 
autoionization-like process has finite lifetime which 
depends on the electric field applied\cite{Kishimoto};  steeper the 
applied electric 
field pulse, shorter time is allowed for the atoms to be ionized with this 
autoionization-like process, thus inducing relative 
increase of the fraction of the ionization component due to the tunneling 
process. 

These observations are the first clear indications of the systematic 
occurrence 
of the tunneling process in these highly excited nonhydrogenic Rydberg atoms 
with $n \geq 90$.  
Previously two groups\cite{Neijzen,Rolfes} observed a field ionization peak
due to the tunneling process for the 66$p$ ($^{2}P$) state in In and 34$d$ 
state in Na.   

%

Finally we comment on the lower peak behavior. 
Contrary to the clear behavior of the higher peak component from the  tunneling 
process, the lower peak can not be explained simply from the 
expected behavior of the autoionization-like process;  if the ionization path 
follows the same diabatic line as in the case of higher peak 
component (tunneling process), the ionization field would depend on the      
excitation position of the  manifold states, approximately 
given by the 
crossing point between the line $F_c$ and each straight line of 
diabatic trajectory in Stark energy  diagram as shown in 
Fig.\ref{fig:112m_tunneling}.  This is not the 
case in the actual situation as shown in the present experiment.  
The behavior is rather consistent with that expected from the 
assumption that the initial manifold states take adiabatic 
paths to ionization at the classical saddle point values $F_c$, although the 
slew rate of the electric field pulse 
applied actually is too high for allowing the adiabatic paths in the 
individual avoided crossings.     
More detailed experimental results together with discussions 
on the characteristic feature of the lower peak component will be 
presented elsewhere.  
%

In conclusion we have studied the pulsed field ionization behavior of highly 
excited Rb Rydberg atoms in high slew rate regime.  Specifically we 
focussed on the field ionization behavior of the manifold states in high $n$ 
region. It was found for 
the first time that the tunneling process plays increasingly the 
dominant role for the field ionization at such highly excited nonhydrogenic 
Rydberg atoms with $n$ 
ranging from 90 to 150.  The fraction of the ionization component 
from the tunneling process increases with increasing slew rate. 
This behavior brings us  an important application of high-lying Rydberg 
atoms to serve as a sensitive microwave single-photon 
detector over the 10-cm wavelength region\cite{Matsuki};  in this 
new method, two low-$\ell$ states which lie below and above the 
adjacent manifold, respectively, are selected as the lower and the upper 
states for 
the Rydberg-atom microwave detector.  The lower state initially prepared 
(the upper state populated by absorbing photons) is then  
transferred to the lowermost (uppermost) state in the adjacent manifold 
adiabatically.   The resulting extreme states in the manifold  are then 
selectively 
ionized with the following pulsed electric field in high slew rate regime.  
The method was demonstrated  to be quite stringent in selectivity  
 for the highly excited Rb Rydberg atoms\cite{Tada}.   

This research was partly supported  by a Grant-in-aid for Specially Promoted 
Research by the Ministry of Education, Science, 
Sports, and Culture, Japan (No. 09102010).  M.T., S.Y., and 
T.H. thank the 
financial support of JSPS, Japan  under the Research Fellowships for the Young 
Scientists. 


%


\end{document}